\documentclass[aps,a4paper,
,nofootinbib]{revtex4}
\usepackage{amssymb,latexsym}
\usepackage{graphicx}
\usepackage{bm}
\usepackage{mathrsfs}
\usepackage{xcolor,color,graphicx,graphics}
\usepackage[all]{xy}
\usepackage{epsfig,subfigure}
\usepackage{latexsym,amssymb,amsmath,amsfonts}
\usepackage[english]{babel}
\usepackage[OT1]{fontenc}
\usepackage[latin1]{inputenc}
\usepackage{makeidx}
\usepackage{hyperref}
\usepackage{color,graphicx,graphics,wrapfig,epsf}
\begin{document}
\newcommand\be{\begin{equation}}
\newcommand\ee{\end{equation}}

\title{Stellar dynamics within the virial theorem: asymptotic small-parameters time expansion of the Ermakov-Lewis-Leach invariant as an infinite series of conservation laws}
\author{
\normalsize
Orchidea Maria Lecian\\
\small
lecian@icra.it\\
ICRA c/o Physics Department, Sapienza University of Rome,\\ Piazzale Aldo Moro, 5- 00185 Rome, Italy;\\
Faculty of Civil and Industrial Engineering, Sapienza University of Rome,\\ Via Eudossiana, 18- 00184 Rome, Italy;\\
Faculty of Information Engineering, Sapienza University of Rome,\\ Via Eudossiana, 18- 00184 Rome, Italy;\\
\normalsize Brunello B. Tirozzi\\
\small brunellotirozzi@gmail.com\\
Department of Physics,\\
\small
University La Sapienza of Rome, Italy.\normalsize
}
\maketitle
\section*{Abstract}
The regime of undamped oscillations characterizing the stellar dynamics of virialised systems is analysed within the framework of a new approach to the study of the integrals of motion. The method is relevant as far as the cosmological implementation is concerned, as it applies to the calculations apt for any age of the evolution of the Universe starting from the epoch of non-Gaussianities until present time. The new method here developed is based on the asymptotic small-parameters expansion of the new expression of the Ermakov-Lewis-Leach integrals of motion; as a result, an infinite series of new conservation laws implies the uniqueness and existence of new integrals of motion; analytical examples are provided after the pulsed Plummer potential, the three instances of the pulsed Dehnen potentials, the pulsed harmonic potential, the Jaffe potential, the Hernquist potential, applied to the studied problem. Particular cases of complex potentials are therefore also comprehended in the analysis. The constants of motions descending from the conservation laws are demonstrated to depend on the virialised radius and on the virialised mass of the stellar system, independently of the potential used.\\
COsmological implemetation is given within the framework of a generic Terzic'-Kandrup potential.\\
Comparison with data analysis techniques are provided with.\\
{\bf Keywords}: Astrophysics dynamics; non-Gaussianities; virialised systems; integrals of motion; dark matter.

\section{Introduction\label{section1}}
\vskip 40 pt
The interest of Astrophysics dynamics of virialised systems relies on the possibility to study the formation of astrophysical objects and their evolution within the virialised description, which can be assumed to be starting at the epoch of non-Gaussianities, and to hold up to present times \cite{nga}, within the framework of the collisionless assumptions.\\
The stellar stability was phenomenologically analysed in \cite{wes39} as far as the period density and the light curves are concerned.\\
Observational evidence of galaxy oscillations is reported in \cite{msm94}.\\
Some features of the asymptotic theory of stellar oscillations and of excitation and damping of the oscillations are recapitulated in \cite{asy}.\\

In \cite{kud15}, the fitting techniques of stellar stability of time-dependent radial movement of mass shell modelised after an oscillator equation relating the eigenfrequency with the adiabatic exponent are presented.\\
One of the advantages of studying the properties of the integral constants of motion for virialised systems is the possibility to trace the wanted properties of the dark-matter effects within stellar systems, as in \cite{Xu:2022ypi}.\\
In \cite{bin78}, the problem is addressed, about the description of galaxies as prolate, oblate or triaxial. The numerical method followed mimics the representation of the density distribution produced after one orbit when averaged over a long time. The quantity $\tilde{E}=\frac{1}{2}v^2+\tilde{Phi}$ is found to be always isolating; nevertheless, the quantities $f(\tilde{E}), L_z$ are commented as not suited as elliptical galaxies have been measured to rotate faster (pag. 29 ibidem).\\
In \cite{ste94}, the relaxation of the assumption that the potential in the Jacobi integral be an even function of the coordinates allows one to further investigate the dynamics of two-dimensional rotating systems.\\
In \cite{sch79}, the features of triaxial galaxies are investigated numerically; a density distribution with a modified Hubble profile is taken to approximate elliptical galaxies. As a result, the orbits computed are found to have three effective integral: the energy and two non-classical integrals. For these purposes, a model density distribution is reproduced after superposing the orbital density distributions if each orbit is reasonably occupied by an appropriate number of states, with occupation numbers being non-negative.\\
In \cite{vra14}, an analytical formula for an approximation of the third integral of motion in some specific case is provided with.
The delicate issue of treating the integration boundaries according to the Standard Cosmological Principle has been addressed only recently and only numerically in \cite{phe20} for the choice of an appropriate scale factor in GR.  The analysis of \cite{gea19} is focused on galaxy clusters. Differently, the studies of \cite{KBR12} point out the setting of the virialising radii in the computation; to do so, the conduction length is demonstrated to be
typically less than the size of the virializing cluster also in the absence of
any suppression of the thermal conductivity for the choice of the conduction length: the choice of a realistic suppression factor is discussed.
The study \cite{phe20} is concerned with choosing a reasonable size to integrate over the sampled Astrophysical system.\\
The realistic implementation of the virializing techniques of self-gravitating (Relativistic) Riemann-like galaxies is envisaged in \cite{vandevoort3}.\\
In \cite{sri89}, the features of undamped oscillations of stellar systems are described; to do so, the Lewis invariant is used.\\
The virialised radius is not always found as an effective fitting parameter from the fitting techniques of the observational evidence \cite{lew02}.\\
A discussion of the use of the virial masses within fitting algorithms is provided with in \cite{tkk08}.\\
In \cite{perlin}, coherent oscillations of the entire stellar system under the effect of a perturbing mass are studied.\\
The problem of visibility of oscillations was recently discussed in \cite{vis}: in comparison with the signal-to-noise ratio, a time series is obtained after realistic magneto-hydro-dynamic simulations, in combination with radiative transfer calculations; the interpretation of the time series is left as an open issue.\\
The purposes of the present paper are the formulation of conservation laws of the constant of motions descending from the establishment of the technique of the asypmtotic small-parameters time expansion, which lead to constants of motion depending on the virial mass and on the virial radius at the cosmological implementation.\\
The technique is applied to the Lewis-Ermakov-Leach invariant as far as the analysis of stellar systems is concerned.\\
In the present work, the constant of motion is expressed as a function of the virialised radius, and as a function of the virial mass of the stellar objects (where the latter result holds due to the non-restrictive assumption that, during an infinitesimal time interval at the time of non-Gaussianities, the mass of the stellar object is not varying, i.e., for instance, there is no ejection, such that the phenomena described in \cite{moe12} and \cite{wei100} are avoided).\\
The results are obtained after the consideration of a stellar objects whose properties are ruled after a generalised potential, which can contain oscillating terms. The integral of motion is decomposed according to the small time-parameters technique, and new conservation laws of new integrals of motions are found.\\
The cosmological implementation is provided with, within the framewrok of a generic Terzic'-Kandrup potential.\\
The methodologies straightfroward apply to the Ermakov-Lewis invariant and to the Ermakov-Lewis adiabatic invariant. 
The system can tested at different approximation orders.\\
The cosmological implementation is achieved at the time of non-Gaussianities. According to this choice, it is newly possible to analyse the oscillation properties of stellar systems as due to two different components, i.e. the constants of motions, and the phenomena perturbing the CBE system at any time of the observational evidence.\\ 
The interest in these new description is also that it is possible to apply the paradigm to calculate the cosntants of motion also at different ages of the Universe, at which stellar systems can be hypothesised to be started forming.\\
The manuscript is organised as follows.\\
In Section \ref{section1}, the main achievements regarding the constants of motion, the virial radius and the virial mass of stellar systems are introduced.\\
In Section \ref{section2}, the application of the tensor version of the virial theorem to stellar objects is recalled to be reconducted to the scalar version, at which there defines a virialised radius.\\
In Section \ref{section3}, the elements that lead to a choice of a collisionless Boltzmann distribution function are discussed.\\
In Section \ref{section4}, the phenomenon of oscillations of stellar systems is revised.\\ 
In Section \ref{section5}, the mathematical tools needed for the definition of generalised Hamiltonians are summarised.\\
in Section \ref{section6}, time-dependent oscillator potentials are recapitulated.\\
In Section \ref{section7}, new estimations of the constants of motions after the Lewis-Ermakov-Leach invariant are found. The method can be analysed as to apply strightforward to the Ermakov Lewis invariant and to the adiabatic Ermakov-Lewis invariant.\\
In Section \ref{cola}, the new conservation laws of the new constants of motions are stated.\\
In Section \ref{section12}, the cosmological implementation of the newly-found conservation laws of the constants of motion is provided at the epoch of non-Gaussianities.\\ 
In Section \ref{modcbe}, the modifications of the distribution function of collisionless-boltzmann systems are discussed as far as the Astrophysical implementation is concerned.
In Section \ref{outlook}, the further issues concerning stationary oscillations are this way freshly analysed.\\
In Section \ref{perspectives}, the phenomena which lead away from the description of stationary oscillations are thus anew envisaged.\\
In Section \ref{remarks}, the guidelines of investigation of the cosmological issues allow one to frame the new results here demonstrated.
\section{Stellar dynamics within the virial theorem\label{section2}}
In \cite{cel72}, after \cite{ced68}, the tensor version of the virial theorem, applied to a system of equal-mass points, after the Liouville equation, is specified to the dynamics of the $6$-dim phase-space of spherical stellar systems. More in detail, stellar systems with positive total energy are demonstrated to disperse at infinity; differently, spherical systems with negative total energy are demonstrated to perform periodic oscillations with finite amplitude. The implementation of the technique to Astrophysical systems spams from clusters of stars to clusters of galaxies.\\
The scalar version of the virial system in the hypothesis of stationarity, after a kinetic energy $\mathcal{T}$ and a potential energy $\mathcal{W}$, is stated as
\begin{equation}
2\mathcal{T}+\mathcal{W}=0;
\end{equation}
the gravitational potential energy $\mathcal{W}$ is written
\begin{equation}
    \mathcal{W}=-\frac{1}{2}\frac{GM^2}{\bar{R}},
\end{equation}
with $\bar{R}$ being the average measure of the linear dimension of the system (i.e., also, the radius of the sphere in a spherical system), and $M$ the mass.\\
Two-points correlations are to be ignored.\\
The following derivation holds
\begin{equation}
\frac{d^2}{dt^2}\mathcal{I}_{ij}=2\mathcal{T}_{ij}+\mathcal{W}_{ij},
\end{equation}
where $\mathcal{I}_{ij}$ is the inertia-moment tensor, $\mathcal{T}_{ij}$ is the kinetic-energy tensor, and $\mathcal{W}_{ij}$ is the potential-energy tensor.\\
Be $\mathcal{E}$ the total energy of the system, remaining constant: the following application holds
\begin{equation}\label{eq6}
  \frac{d^2I}{dt^2}=2\mathcal{T}+\mathcal{W}=2\mathcal{E}-\mathcal{W}.  
\end{equation}
It is therefore now possible to analyse the evolution of the dynamics of the studied spherical system.\\
The dynamics is hypothesized to start at some initial time, at which a cluster is characterised after its moment of inertia, its kinetic energy and its potential energy. The evolution is therefore described after Eq. (\ref{eq6}).\\
It is momentous to recall the attention on the choice of the linear size  $\bar{R}$, i.e. the internal distribution of the density cluster is demonstrated not to be of relevance. For these purposes, one chooses s spherically-symmetric cluster consisting of $N$ equal-mass points, distributed in the phase space according to a Gaussian density distribution.\\
Therefore, the density $\tilde{\rho(r)}$ reads
\begin{equation}\label{eq22}
\tilde{\rho(r)}=\frac{Nm}{\mathcal{A}^3\pi^{3/2}}e^{-r^2/d^2},
\end{equation}
with $\mathcal{A}\equiv\mathcal{A}(t)$ the radius of the cluster.\\
Eq. (\ref{eq22}) is assumed to be maintained.\\
Eq. (\ref{eq6}) implies that
\begin{equation}\label{eq23}
\frac{3}{4} Nm\frac{d^2\mathcal{A}}{dt^2}=2\mathcal{E}+\frac{GN^2m^2}{(2\pi)^{1/2}}.
\end{equation}
\subsection{Virialization}
It is relevant to remark that, if instead of a Gaussian distribution, a homogeneous distribution had been chosen inside the space of radius $\mathcal{A}$, an equivalent equation, differing only after the coefficients, would have been obtained, which will assume the same dimensionless form as follows.\\
{\bf Def.}: Be $\mathcal{A}\equiv\mathcal{A_0}z$, with the specification of $\mathcal{A_0}$ to be designated 
 and measuring the time VEDI in
\begin{equation}\label{eq25}
t_0=\left(\frac{9}{8}\pi\right)^\frac{1}{2}\frac{\mathcal{A_0}}{NmG}.
\end{equation}
Eq. (\ref{eq23}) implies the role of $\mathcal{E}$ to be spelled out following
\begin{equation}\label{eq26}
\frac{d^2z}{dt^2}=2\left(sign\mathcal{E}\right)+\frac{1}{z}
\end{equation}
with the introduction of $Q$ as
\begin{equation}\label{eq27}
Q=\left(\frac{\mathcal{E}}{\mathcal{W}_0}\right)
\end{equation}
being $\mathcal{W_0}$ the potential energy at $\mathcal{A_0}$.\\
Finally, Eq. (\ref{eq26}) admits first integral
\begin{equation}\label{eq28}
z^2\left(\frac{dz}{dt}\right)^2=\left(sign(\mathcal{E})\right)Qz^2+z+const
\end{equation}
\subsubsection{The application to systems wit negative total energy}
For systems with negative total energy, Eq. (\ref{eq28}) rewrites with the help of the integration constant $\Lambda$
\begin{equation}\label{eq33}
z^2\left(\frac{dz}{dt}\right)^2=\Lambda^2-Q\left(z-\frac{1}{2Q}\right)^2.
\end{equation}
Eq. (\ref{eq33}) is solved for $t$ 
\begin{equation}\label{eq34}
t=\frac{1}{Q^{1/2}}\left[\Lambda^2-\left(z-\frac{1}{2Q}\right)\right]^{1/2}+\frac{1}{2Q^{3/2}}cos^{-1}\left[\frac{1}{\Lambda}\left(z-\frac{1}{2Q}\right)\right].
\end{equation}
 It is possible to define $\mathcal{C'}$ an arbitrary constant defining the origin of $t$; the constant $\mathcal{C'}$ is unimportant and can be neglected.\\
Eq. (\ref{eq34}) implied that, when the total energy of the system is negative, the system performs periodic oscillations of finite amplitude, characterized after a period $p$
\begin{equation}\label{eq35}
    p=\frac{\pi}{Q^{3/2}}
\end{equation}
and amplitude of oscillation for $z$ as
\begin{equation}\label{eq36}
    \Lambda+\frac{1}{2Q}\ge z\ge-\Lambda+\frac{1}{2Q}.
\end{equation}
The characterizing case $z\ge 0$ is, by definition,
\begin{equation}
    \Lambda\le\frac{1}{2Q}.
\end{equation}
The special case
\begin{equation}
    \Lambda=\frac{1}{2Q},
\end{equation}
i.e. vanishing kinetic energy at $t=0$ and $t=\frac{1}{Q}$.
 \section{Thee choice of the collisionless Boltzmann distribution in Astrophysical systems\label{section3}}
The problem of a large number $N$ of interacting particles is here addressed after the consideration of a small number of quantities, which are averaged in a specific manner, rather than that of the exact position $\vec{r}$ and velocity $\vec{v}$ of each particle \cite{fpo84}. The $N$ interacting particles in large number can be described as a function of the number $N$, the density and the collision frequency at characteristic times and spacial scales.\\
\paragraph{The Vlasov equations}
It is necessary to reconcile the Newtonian equations of motion with those of hydrodynamics within the description of motion in continuous media.\\
A hydrodynamical systems is characterised at a given point after the density $\tilde{\rho(\vec{r}, t)}$, the pressure $P(\vec{r}, t)$ and the velocity of motion of the fluid $\vec{v}(\vec{r}, t)$.\\
There therefore holds the continuity equation
\begin{equation}\label{fp1}
\frac{\partial \tilde{\rho}}{\partial t}+div(\tilde{\rho} \vec{v})=0
\end{equation} 
to be combined with the Euler equations
\begin{equation}\label{fp2}
\frac{d\vec{v}}{dt}\equiv \frac{\partial \vec{v}}{\partial t}+(\vec{v}\nabla)\vec{v}=-\frac{1}{\tilde{\rho}}\frac{\partial P}{\partial \vec{r}}+\vec{F},
\end{equation}
$\vec{F}$ being the force per unit mass.\\
In view of the requested reconciliation, the force per unit mass $\vec{F}$ has to be set to imply the Newtonian gravitational force
\begin{equation}
\vec{F}=-grad \Phi(\vec{r}, t),
\end{equation}
being $\Phi(\vec{r}, t)$ the gravitational potential.\\
Therefore, combining the continuity equation Eq. (\ref{fp1}) and the Euler equations Eq.'s (\ref{fp2}), the hydrodynamical description is achieved after posing $\vec{F}=0$ in a chosen equation of state. Differently, if $\vec{F}\neq0$, in the case of gravitating systems, the Poisson equation
\begin{equation}\label{fp6}
\Delta\Phi=4\pi G\tilde{\rho}
\end{equation} 
has to be recovered.\\
The thermal motion of particles is schematised in the kinetic theory, in which the statistical description occupies a central role: the distribution function $f(\vec{r}, \vec{v}, t)$ has to be set after the particles.\\
The Boltzmann kinetic equation reads
\begin{equation}\label{fpa}
\frac{df}{dt}=\mathcal{C}
\end{equation}
along the exact trajectory of the particle in the field of the force form which $\vec{F}$ is normalised.\\
The choice $\mathcal{C}=0$ in Eq. (\ref{fpa}) corresponds to a collisionless kinetic equation.\\
The choice $\mathcal{C}=0$ adapts obviously to Astrophysical systems. It corresponds to he incompressible feature of a phase fluid, after which the distribution function stays unchanged, i.e.
\begin{equation}\label{fp9}
\frac{df}{dt}=0
\end{equation}
during the path of the phase space.\\
The density $\tilde{\rho}(t)$ therefore reads
\begin{equation}\label{fp10}
\tilde{\rho}(\vec{r}, t)=\int f(\vec{r}, \vec{v}, t) d\vec{v}.
\end{equation}
The set of equations Eq. (\ref{fp9}), Eq. (\ref{fp6}), and Eq. (\ref{fp10}) is denominated Vlasov equations.\\
The set of Vlasov equations is analogous to the equations characterising a plasma.
\paragraph{The collisionless kinetic equation}
The collisionless kinetic equation Eq. (\ref{fp9}) describes the evolution of the single-particle distribution function $f=f(\vec{r}, \vec{v}, t)$. It can be obtained also from the Liouville equation of the $N$-particle distribution function $\mathcal{F}=\mathcal{F}(\vec{r_1}, \vec{v_1}, \vec{r_2}, \vec{v_2}, ...; t)$ which is a function of the particles coordinates, the particles velocities and of time, as
\begin{equation}\label{fpb}
\frac{\partial \mathcal{F}}{\partial t}+[\mathcal{F}, H]=0,
\end{equation}
 endowed with its Poisson brackets, being $H$ the Hamiltonian of the function (where the Hamiltonian variables $q_\alpha$ and $p_\alpha$ are defined).\\
Eq. (\ref{fpb}) rewrites also
\begin{equation}\label{fp15}
\frac{\partial \mathcal{F}}{\mathcal t}+\sum_{\alpha}v_\alpha\frac{\partial \mathcal{F}}{\partial x_\alpha}+\sum_{\alpha}\mathcal{F}_\alpha\frac{\partial \mathcal{F}}{\partial v_\alpha}=0.
\end{equation}
Eq. (\ref{fp9}) is therefore also obtained from Eq. (\ref{fp15}) after integration over the portion of the phase space which is comprehended within the phase-space coordinates of the particles except for those of one particle; the function $\mathcal{F}$ therefore splits as
\begin{equation}\label{fp16}
\mathcal{F}=f(\vec{r_1}, \vec{v_1}, t)f(\vec{r_2}, \vec{v_2}, t)...f(\vec{r_N}, \vec{v_N}, t).
\end{equation} 
Eq. (\ref{fp16}) is therefore commented as suited for Astrophysical systems, where the statistical independence of the phase distributions of different objects is requested, and in which the collisions (i.e. the interaction forces between the individual pairs of particles) are negligibly small with respect to the 'smoothed' force $\vec{F}$. The smoothed force $\vec{F}$ here considered is one due to the collective self-consistent action of all the particles of the system.\\
The self-consistent nature of the field is due to the fact that the field is produced after a particle distribution $f$, which, on it turn, is produced after the same field.\\
From  VEDI [86, 138], one notes that the condition of collisions to be neglected is reflected int he requirement that the number of particles in the so-called 'Debye-sphere' is large.\\
\subparagraph{Collisionless gravitating system and the equilibrium states} The equilibrium states of a collisionless gravitating system are described after the distribution functions $f_0(\vec{r}, \vec{v})$, the mass density $\tilde{\rho}_0(\vec{r})$, and the gravitational potential $\Phi_0(\vec{r})$. They obey the Vlasov equations Eq. (\ref{fp6}), Eq. (\ref{fp9}), Eq. (\ref{fp10}) where $\frac{\partial}{\partial t}=0$ is hypothesized.\\
\subsection{Small oscillations of gravitating systems: estimates of the equilibrium configurations}
The gravitating systems are parameterised after a distribution function $f_0(\vec{r}, \vec{v})$, a mass density $\tilde{\rho}_0(\vec{r})$, related after a gravitational potential $\tilde{\rho}_0(r)$ which obey the simplified Vlasov equations with $\frac{\partial}{\partial t}=0$ \cite{fpo84}.\\
They have to bey the simplified Vlasov equations as
\begin{equation}\label{fp18}
\vec{v}\frac{\partial f_0}{\partial \vec{r}}-\frac{\partial \Phi_0}{\partial \vec{r}}\frac{\partial f_0}{\partial \vec{v}}=0,
\end{equation}
\begin{equation}\label{fp19}
\Delta\Phi_0=4\pi G\tilde{\rho}_0,
\end{equation}
and
\begin{equation}\label{fp20}
\tilde{\rho}_0=\int f_0d\vec{v}.
\end{equation}
The kinetic equation Eq. (\ref{fp18}) is commented as an equation of the distribution function $f_0$ after having assumed the potential $\Phi_0$ to be known; it is therefore a homogenous differential equation of the first order at partial derivatives.\\
The equations of motion of a particle, ruled after the potential from t he gravitational force $-\frac{\partial \Phi_0}{\partial r}$ are the characteristics of the equations
\begin{subequations}\label{fp21}
\begin{align}
&\frac{d\vec{r}}{dt}=\vec{v},\\
&\frac{d\vec{v}}{dt}=-\frac{\partial \Phi_0}{\partial \vec{r}}.
\end{align}
\end{subequations}
A general solution of Eq. (\ref{fp18}) is an arbitrary function of the particle integrals of motion in the field $\Phi_0(\vec{r})$ [VEDI 156].\\
The requirement of unambiguity of the distribution function for all the points of the phase space leads to the request, on its turn, that the integrals of motion, which can be arguments of the distribution, to be single-valued.\\
The equations of the characteristics of the system of Eq.'s (\ref{fp21}) in Cartesian coordinates write\\
VEDERE MISPRINT LIBRO
\begin{equation}\label{fp22}
\frac{dx}{dv_x}=\frac{dy}{dv_y}=\frac{dz}{dv_z}=-\frac{dv_y}{-\partial\Phi_0/\partial y}=-\frac{dv_z}{-\partial\Phi_0/\partial z}=dt.
\end{equation} 
From Eq. (\ref{fp22}), six independent integral of motion are given, five of which do not dependent on the time $t$.\\
The requirement that the integrals be single-valued has to be compared with the properties of uniformity of time (energy), as well an with those of homogeneity and isotropy of space.\\
Thus, the particle energy $E$ in the field $\Phi_0$ can all ways be one of the arguments of the function $f_0$ as
\begin{equation}
E=\frac{v^2}{2}+\Phi_0.
\end{equation}

\section{About oscillations\label{section4}}
The relations between the isolating integrals, the Jeans theorem and the distribution functions are enumerated in \cite{rug}.\\
In \cite{sri89}, the performance of undamped oscillations of stellar systems is analysed from the definition of the role of the integrals of motions, after \cite{fpo84}.\\
The dynamics of collisionless Boltzmann equations (CBE) is chosen for the schematization of the system as one from the distribution function
\begin{equation}\label{eqa}
\frac{df}{dt}+\vec{v}\cdot\frac{df}{d\vec{r}}-\frac{d\Phi}{d\vec{r}}+\frac{df}{d\vec{v}}=0,
\end{equation}
which implies convective features in the dynamics of stars in galaxies, where the former is ruled after the CBE: in Eq. \ref{eqa}, the distribution  $f$ is defined as $f\equiv f(\vec{r}, \vec{v}, t)$, i.e. such that $f(\vec{r}, \vec{v}, t)d^3rd^3v$ is the mass comprehended in portion $d^3r, d^3v$ of the phase space available for the system at the defined time $t$; furthermore, the potential $\Phi$ is $\Phi\equiv\Phi(\vec{r}, t)$ the gravitational potential compatible with the (self-consistent) Poisson equation
\begin{equation}
\nabla^2\Phi=4\pi G\tilde{\rho}(\vec{r}, t)\equiv 4\pi G\int f(\vec{r}, \vec{v}, t)d^3v
\end{equation}
of the density $\tilde{\rho}\equiv \tilde{\rho}(\vec{r}, t)$.\\As a result, most solutions can be demonstrated to relax into a time-independent description.\\
The standard Cosmological principle can be adopted in order to perform the requested integrals; according to application of the Standard Cosmological Principle, the integration boundaries are chosen.This way, the change of the shape and that of the size of the region is analysed to change in time (while the region is assumed to maintain its geometrical features, i.e. such as ellipsoidal): the strength of the potential is in this manner time-dependent.\\
\subsection{One-dimensional models} One-dimensional modes are found to be characterizes such as all solutions are periodic functions of time.\\Within one-dimensional models, spherical models and elliptical models can be studied.
\subsection{Three-dimensional spherical models} within the three-dimensional models, the system of a homogenous static sphere corresponds to the characterization of a polytropic equation of index zero. Nevertheless, the former schematization is not consistent with the request of isotropic dispersion relation \cite{van80}.\\
A distribution function which depends only on energy is calculated not to describe a uniform sphere; as an an alternative derivation, given $\vec{L}=\vec{r}\wedge\vec{v}$, the distribution function is demonstrated to be requested to depend both on $E$ and on $L$.\\
The inversion of the integral equation of $f$ is studied not to be unique.\\
\section{The time-dependent harmonic oscillator\label{section5}}
The dynamics of the time-dependent harmonic oscillation is ruled after the potential characterizing the Hamiltonian
\begin{equation}\label{tdho}
H=\frac{1}{2\eta}\left[p^2+\Omega^2(t)q^2\right],
\end{equation}
for which there exists a class of exact invariants.\\
The class of exact invariants $I_\eta(\rho)$, where $\rho$ is an auxiliary variable, can be given in a closed form as a function of $\rho(t)$, i.e.
\begin{equation}\label{eqrho}
I_\eta(\rho):\ \ \eta^2\frac{d^2\rho}{dt^2}+\Omega^2(t)\rho-\rho^{-3}=0.
\end{equation}
On its turn, the density $\rho(t)$ obeys the equation SCRIVI, i.e. such that, for each for each particular solution of the equation of $\rho$, an invariant is defined.\\Such an analysis lead to results more general than the asymptotic treatment 
(also in the case of a complex $\Omega(t)$).
\subsection{More about the CBE}
In \cite{lge88}, quadratic time-dependent one-dimensional models are constructed numerically, i.e. in an extended phase space after the Jeans Theorem. In particular, the numerical analysis of spherical oscillations is described to be difficult to perform. By contrast, the presence of non-linear periodically time-dependent solutions of the Poisson equations and the CBE are shown to be possible. The application to bar galaxies is suggested. This suggestion has to be compared with the analytical studies of \cite{vandevoort1} and \cite{van80}, in which, nevertheless, a different distribution function is chosen, under the Chandrasekhar's guideline it should be of a negative energy.\\
In the case of quadratic time-dependent potentials, all solutions are found as periodic function of time $t$; one-parameter of families of oscillating models, with a specified total mass and with a specified energy, are defined. More precisely, it is possible to spell the chose parameter as the first integral.
\subsection{The Lewis invariant}
The Lewis invariant is obtained from Eq. (\ref{tdho}) for complex potential term $\Omega$.
\paragraph{Derivation of the Lewis invariant}
The Lewis invariant is derived in \cite{lew68} in the case of a complex potential $\Omega$ after following the prescriptions of the analysis of \cite{kru62} in the case of a real potential.\\ 
In particular, in \cite{kru62}, the the analysis of Hamiltonian systems whose solutions are all nearly periodic is developed. To this purposes, autonomous systems are newly viewed, and the recurrent systems are accordingly classified, of which the asymptotical solution is found; splittable systems are studied of which the most direct series solution is shown to be not adequate. The standardisation procedure is therefore provided with: appropriate variables are chosen, after which a recursive construction of the requested functions is built.\\
In \cite{lew68}, the procedure followed is dictated after the treatment of \cite{kru62} as implemented in the case of a complex potential $\Omega(t)$ form the Hamiltonian
\be
H=\frac{1}{2\eta} \left(p^2+\Omega^2(t)q^2\right)
\ee 
being $q$ the canonical coordinate, $p$ the canonical conjugate momenta, $\Omega(t)$ an arbitrary complex function of the time $t$, and $\eta$ a positive real parameter.\\
There exists a class of exact invariants
\be\label{lew3}
I=\frac{1}{2}\left[\rho^{-2}q^2+\left(\rho p-\eta \frac{d\rho}{dt}q \right)^2\right]
\ee
with $\rho$ satisfying
\be\label{lew4}
\eta^2\frac{d^\rho}{dt^2}+\Omega(t)^2\rho-\rho^3=0
\ee
The class of exact invariants is therefore defined for any $\rho$ obeying Eq. (\ref{lew4}).\\
If Eq. (\ref{lew4}) can be solved recursively, the series describing $\rho$ is a series of positive powers of the parameter $\eta$; this implies that Eq. (\ref{lew3}) is a series of positive powers of $\eta$.
\paragraph{Implementation of the Lewis invariant}As from the result \cite{lew68}, the following expressions of integrals $I_\rho$ is found.\\
From the comparison with \cite{kru62}, in classical systems with real $\Omega$ $I$ is the series of usual adiabatic-invariant series in which the leading term is proportional to $\eta H/\Omega$.\\
In the present case
\begin{equation}
I_\rho\equiv\frac{1}{2}\left[\rho^{-2}x^2+\rho p-\dot{\rho}x^2\right],
\end{equation}
i.e. one for each of the $\rho$ satisfying the condition
\begin{equation}
\ddot{\rho}+\Omega^2(t)\rho=\rho^{-3}.
\end{equation}
The methods of \cite{kru62} are applied in \cite{lew68} to any complex $\Omega$, $q$ and $p$ at any particular solution of Eq. (\ref{lew4}) after differentiating Eq. (\ref{lew3}), using the Hamiltonian equations to eliminate the first time derivative of the canonical variable and that of hte canonical conjugate momenta, and taking advantage of Eq. (\ref{lew4}) to eliminate the second time derivative of $\rho$.\\
The integral is found in Eq. (34) of \cite{lew68} and represented in terms of $\rho(t)$; the integral is found after Eq. (15a) of \cite{lew68} from the $I=\int p\cdot dq$ on particular closed curves (named rings) after the choice of opportune variables.\\
The rings are demonstrated to be ellipses in the opportune coordinates, after which the first integral can be reversed as a function the Hamiltonian variables $p$ and $q$.
Correspondingly, there exists a conserved symmetric tensor
$I_{mn}$\\
\begin{equation}
    I_{mn}=\frac{1}{2}\rho^{-2}q_mq_n+(\rho p_m-\dot{\rho}q_m)(\rho p_n-\dot{\rho}q_n)
\end{equation}
which is a representation of $SU(3)$.

\subsection{More about invariants}In \cite{lea78}, a generalization of the Lewis invariants is proposed, for the time-dependent oscillator
\begin{equation}
    H=\frac{1}{2}p^2+\frac{1}{2}\omega^2(t)x^2,
\end{equation}
where the potential ruling the dynamics is specified.\\
In \cite{gle}, the work of \cite{lew68} was applied to three-dimensional time-dependent oscillators, which problem comprehends also the anisotropic perturbation problems and the singular quadratic perturbation problem.\\
A conserved symmetric tensor operator for the isotropic-oscillator problem is written
\begin{equation}
    I_{mn}=\frac{1}{2}\left[\rho^{-2}q_mq_n+(\rho p_m-\dot{rho}q_m)(\rho p_n-\dot{\rho}q_n)\right],
\end{equation}
whose symmetry group is a 'non-invariance symmetry group' of the three-dimensional time-dependent isotropic oscillator.\\
More in detail, in \cite{lea78}, the only non-trivial quadratic invariant $I$ of the one-dimensional oscillator problem is found as
\begin{subequations}
\begin{align}
&2I=(\rho^{-1}e^{F/2}q)\left(C_1^2+C_2^2+2C_1C_2 cos(2W)\right)+\\
&\left\{(\rho e^{-F/2}p)-(\dot{\rho}-\frac{1}{2}\rho f)e^{F/2}q \right\}^2(c_1^2+C_2^2-2C_1C_2 cos(2W))+\\
&-4(\rho^{-1}e^{-F/2})\left\{\rho e^{-F/2}p-\left(\dot{\rho}-\frac{1}{2}\rho f\right)e^{F/2}q\right\}C_1C_2 sin(2W)
\end{align}
\end{subequations}
where $W$ obeys the definition
\begin{equation}
W\equiv\int_{t_0}^t\rho^{-2}dt:
\end{equation}
 the coordinate transformation needed to obtain the invariants is provided after the matrix which is symplectic under the condition
\begin{equation}
    C_1^2-C_2^2=1.
\end{equation}
Keeping the transformation symplectic assures the fundamental symplectic form to stay unchanged; this way, the phase space available for the model is kept unchanged.\\
The invariant of the undamped oscillations is found after posing $F=0$ with general values of the constants $C_1$ and $C_2$ as
\begin{subequations}\label{lea38}
\begin{align}
&I=\frac{1}{2}\{ \rho^{-2}q^2\left(C_1^2+C_2^2+2C_1C_2cos(2W)\right)+\\
&+(\rho p-\dot{\rho}q)^2\left(C_1^2+C_2^2-2C_1C_2cos(2W)\right)-4\rho^{-1}q(\rho p-\dot{\rho}q)C_1C_2sin(2W)
\}
\end{align}
\end{subequations}
The Lewis invariant is obtained after the choice $C=0$ of
\begin{subequations}
\begin{align}
&C_1=cosh C,\\
&C_2=sinh C.
\end{align}
\end{subequations}
The choice $C=0$ is not arbitrary; differently, it descends from the request that the coordinate transformation stays symplectic, for which the $C$ constant is interpreted as a scaling factor.

\section{About time-dependent oscillator potentials\label{section6}}
Among the several possibilities of modified Hamiltonians, included those whose potential is generalised in ways as to contain time-dependent oscillator potentials, those of interest within the cosmological implementation are here briefly recapitulated. 
\subsection{Mathematical outline}
The regularity assumptions about the time-dependent oscillators potentials were very recently recapitulated in \cite{fio22}.\\
In \cite{lls79}, the features of a generalized time-dependent oscillator potential are spelled out within the Hamiltonian analysis as
\begin{equation}
H=f(t)\frac{p^2}{2m}+\frac{1}{2}g(t)w_0^2x^2
\end{equation}
being $H$ the Hamiltonian a system of a point particle of mass $m$, i.e. for a system in which both the kinetic term and the potential term are generalised as time-dependent by means of the functions $f(t)$ and $g(t)$, respectively. The initial values for the two functions $f(t)$ and $g(t)$ are established.
\subsection{Physical implementations of the time-dependent potentials}

The need of a oscillator symmetry in the description of galactic dynamics is recalled in \cite{ros95}. There, the galactic velocity field is studied as a linear function of the cartesian coordinates of the masses. The definition of collective kinetic energy (as the kinetic energy  within the linear-velocity-field approximation) is explained as one neglecting the degrees of freedom which are connected with non-linear velocity fields: the possibility to render the description of galactic systems is outlined, according to which the theory of symplectic dynamical symmetry is established as far as classical systems are concerned. As a result, the groups, of which the co-adjoint orbit is representing the classical phase space, is set: the symmetries are studied. One of the elements of the symplectic Lie algebra is the matrix composed from the self-gravitating potential energy, the angular velocity and the hydrostatic pressure. The solutions of the Hamiltonian dynamical systems characterised after these symmetries are studied as isospectral deformations. The conserved quantities are found as the Casimirs.\\
In \cite{tca2004}, the case of spherically-symmetric potentials which lead to a periodic dynamics are recalled.\\
There, the potentials proposed $V=V(r, t)$ are of the simplified form
\begin{equation}
V(r, t)=V_(r)(1+m_0 sin \omega t)
\end{equation}
which can be cast in the Mathieu Equation,
because the equation
\begin{equation}
\frac{d^2X}{dt^2}+\Omega^2(t)X=0
\end{equation}
with
\begin{equation}
\Omega^2(t)=a+b cos \omega t
\end{equation}
is recast as
\begin{equation}
\frac{d^2X}{d\tau^2}+(\alpha+\beta cos 2\tau)=0.
\end{equation}
The simplified case $m_0=const$ is relevant in considering the mass(es) as constant. i.e. in comparison with the less simplified cases.\\
Four potentials are studied: the pulsed Plummer potential and three cases of the pulsed Dehnen potential. More in detail, the pulsed Plummer potential reads
\begin{equation}\label{plum}
V(r, t)=-\frac{m(t)}{\sqrt{1+r^2}};
\end{equation}
the three pulsed Dehnen potentials read
\begin{equation}\label{deh}
V(r, t)=-\frac{m(t)}{2-\gamma}\left[1-\frac{r^{2-\gamma}}{(1+r)^{2-\gamma}}\right]
\end{equation}
and are specified after the relevant cases $\gamma=0$, $\gamma=1/2$ and $\gamma=1$. In Eq. (\ref{plum}) and in Eq.'s (\ref{deh}), the mass $m(t)$ can be specified as
\begin{equation}\label{mt}
m(t)=1+m_0 sin \omega t.
\end{equation} 
\paragraph{More about variable galactic mass}
According to these specifications, \cite{gas07}, the periodic motion of a star under particular conditions is investigated under a generalized mass-variation law, and within the Poincar\'e small-parameter method.\\
The generalized law of the galaxy mass variation considered is 
\begin{equation}
M(t)=M_0M^n(t),
\end{equation}
which is also named the analogue of the Eddington-
Jeans law after \cite{jea24} and \cite{polyach89}.\\
\\
The Jaffe potential \cite{jaf83} is a potential which was derived in a way such that the gravitational potential and the projected velocity dispersion are easily outlined.\\
The Hernquist potential \cite{her90} as well enjoys the properties that the analytical expression of the velocity dispersion terms of elementary functions holds.\\
The pulsed Harmonic potential in the case of a classical particle was studied in \cite{nzi98}.\\
The Plummer potential and the Hernquist potential were used to study the mass variation of galactic environments \cite{nrl11}.
In \cite{lea83a}, the harmonic-oscillator potential with mass as a time-dependent function included both in the kinetic term and in the potential term is presented.\\
The Jaffe potential and the Hernquist potential can be considered as particular cases of the Dehnen potential.\\

\paragraph{More about perturbed potentials}
The isotropic-harmonic-oscillator potential perturbed by a polynomial term was studied in \cite{bgl16} and the references therein.

\subsubsection{The Lewis Hamiltonian} In \cite{lew67a} and \cite{lew67b}, the Hamiltonian was considered
\begin{equation}\label{eqh}
H=\frac{1}{2\eta}\left[\frac{p^2}{2m}+\Omega^2(t)q^2\right]
\end{equation}
for which the assumption of $\Omega(t)$ an arbitrary continuous function of time is requested, being $\eta$ a positive, real parameter; the existence of the Hamiltonian $H$ Eq.(\ref{eqh}) is studied.\\
In \cite{mmg20}, Eq. (\ref{eqh}) is further investigated.
The work of \cite{sri89} was generalised to a planar galaxy model in \cite{sri90}, for which a $10$-dim phase space is found.
\subsection{Possible generalizetions: homogenous power-law potentials}
Homogenwous power law time-dependent potentials, which evidently contain the time-dependent harmonic oscillator as a simplified version, are analysed in \cite{rob16}; in particular, the WKB methods are exposed.\\
The mechanisms undrelying variation of mass of stellar structures, i.e. such as revised in \cite{nvi02}, can be in the first analyses disregarded.
\section{Estimations of the constants of motions\label{section7}} 
One poses $\rho\equiv\rho(t)$ as
\begin{equation}\label{eqrhoexpan}
\rho\equiv\rho_0+\epsilon\rho_1+\epsilon^2\rho_2+\epsilon^3\rho_3+...
\end{equation}
with $\epsilon<<1$.

\paragraph{Further approaches}
A method to find exact invariant for the time-dependent harmonic oscillator is discussed in \cite{lea1977}.\\
The generalised time-dependent Hamiltonian is studied in \cite{lea76}.\\
In \cite{sar77}, new transformations are considered.\\
In \cite{lle82}, a class of non-linear, time-dependent potentials is analysed.\\
In \cite{egr76}, the Lewis invariant is studied as projection of an auxiliary two-dimensional motion; the analysis is thus useful for the analysis of the phase space, as recalled throughout the paper.\\
In \cite{and81}, the physical characterization of a cluster of particles with collision processes is considered; the analysis will be of interest in the discussion of \cite{ros95}.\\
In \cite{sri2001}, the invariant considered is as an integral of an
energy-balance equation.



\section{Conservation laws and integrals of motion\label{cola}}
When the expansion of $\rho(t)$ as Eq. (\ref{eqrhoexpan}) is considered at all the orders, the infinite series of conservation laws \cite{polyan} is found as
\begin{equation}\label{eqcola}
\frac{d}{dt}\dot{\rho}^2_{2n}=(-1)^{-1+n/2}\dot{\rho}^2_{2n}f_{2n}(\rho_{2n}, \rho_{2n-2}, ..., \rho_2, \rho_0)
\end{equation}
which defines an infinite series of integrals of motion $\rho_{2n}$.\\
Indeed, Eq. (\ref{eqcola}) is demonstrated to be not apt to be written as an Abel equation, nor as an Emden-Fowler-like equation.
\section{Cosmological implementation: conserved integrals at the time of non-Gaussianities \label{section12}}
It is the aim of the present Section to provide with a cosmological implementation of the analytical results here obtained. As a previous remark, one should notice that, one one hand, the fit analyses of the experimental evidence at present times does not agree with the specification of the virialised radius mathematically expected; on the other hand, the presence of elements modifying the specifications of the CBE assumptions are very strong, as further specified in Section \ref{modcbe}.\\
It is therefore the opportunity to state the exact conservation laws of the integrals of motion and to test them at the very beginning of the objects formations, during which (integration) lapse of time the further effects can be neglected; the obtained results are therefore to be compared with the then-present further cosmological ingredients.\\
For these purposes, it is our goal to study the solutions of the systems implemented in Section \ref{cola} at the very beginning of the epoch of non-Guassianities, during a small time interval when even $\tilde{\rho}_2$ can be considered as very-slowly varying, and during during which all the other modifications to the CBE-hypothesis are considered as not effective.\\
As a result, it is thus possible to understand the mechanisms that lead the objects formations after the exact integrals of motions obtained after the conservation laws are established, i.e. the effects modifying the CBE hypothesis, for which, as an examples, the estimations of the virialised radii of objects does not fulfill the fitting algorithms. For the latter reason, the component $F(r)$ of the potentials can be considered as slowly-varying, i.e. during the considered time integral the virialised radius $r$ is considered as the mathematical virialised one $r_v$.\\
During this very small time interval $\Delta t\equiv t_f-t_i$, the conserved component $\rho_0$ is evaluated after a series expansion in the time variable as
\begin{equation}
\int_{t_i}^{t_f}\int_{t_i}^t\ddot{\rho}_0(\tau)\simeq A_0\frac{t_f^2}{2}-A_0\frac{t_i^2}{2}-A_0t_it_f+A_0\frac{t_i^2}{2}+A_1\frac{t_f^3}{6}-A_1\frac{t_i^3}{6}-A_1\frac{t_it_f}{2}+A_1\frac{t_i^2}{2}+A_2\frac{t_f^4}{12}-A_2\frac{t_i^4}{12}-A_2\frac{t_i^3t_f}{3}+A_2\frac{t_i^4}{3}
\end{equation}
with the constants
\begin{subequations}
\begin{align}
&A_0\equiv\frac{4\tilde{\rho}_2}{F(r_v)}\frac{arctanh\sqrt{\frac{m_0}{m_0^2-1}}}{\omega\sqrt{m_0^2-1}},\\
&A_1\equiv\frac{2\tilde{\rho}_2}{F(r_v)}\frac{1}{(m_0^2-1)(1-\frac{1}{m_0^2-1})},\\
&A_3\equiv-\frac{\tilde{\rho}_2}{F(r_v)}\frac{\omega m_0}{(m_0^2-1)(1-\frac{m_0^2}{m_0^2-1})}.
\end{align}
\end{subequations}
The dependence of the constant of motion $\rho_0$ on the frequency $\omega$, on the mass $m_0$ and on the radius of the simplified pulsed potentials is therefore delineated. Moreover, the constant of motion is expressed as a function of the virialised radius. As a result, the mass $m_0$ entering the simplied definitions of the pulsed potentials is here demonstrated to be therefore the virialised mass.
\subsection{The constants of motions as function of the viralised radius and of the virialised mass}
According to the cosmological implementation achieved at the age of the Universe of non-Gaussianities, it is possible to analyse the dynamics of stellar systems as due to different contributions: one one hand, the constants of motions are demonstrated to be functions of the viralised rardius and of the virialised mass within the framework of the CBE choice; on the other hand, the modifications with respoect to the CBE dynamics is due to the different phenomena characterising the evolution of the stellar sysyem. 
\section{Modifications of the CBE hypothesis \label{modcbe}}
The scheme of the linear approximation of the perturbation to the CBE after a point mass is studied in \cite{perlin}. The further generalisations of the simplified scheme lead to the analyses of \cite{ros95}.
\subsection{Adiabatic invariants}
The origins of the adiabatic invariant in classical mechanics are traced in \cite{adi1}.\\
Some indirect relations between the change in action of the harmonic oscillator and slowly-varying perturbations are analysed in \cite{wei94}.\\
In \cite{bpu}, a broad overview about the use of adiabatic invariants is presented.\\
In \cite{smi10}, the adiabatic invariant is used to calculate the eccentricity of the orbit of a globular cluster in a particular self-gravitating system.\\
As a toymodel example, in \cite{syo96}, the one-dimensional harmonic oscillator with a slowly-varying frequency is considered; as an example, a
one-particle system in a slowly varying isochrone potential is described. The analysis of adiabatic invariants demonstrates not to produce secular errors in the computation after symplectic integrators.
\subsection{About 'irregular' forces}
The Kuzmin integral is hypothesized to be a function depending on the velocities only \cite{kuz53}, \cite{kuz56}. The triaxial velocities distribution is studied in \cite{kuz87}. Irregular forces in the secular evolution of a stellar system are studied in \cite{kuz63}. The radial velocities dispersions for comparison with the Kuzmin formulation is studied in \cite{04845}.\\
More in detail, in \cite{kuz53}, stationary galaxies are analysed as far as the definition of integrals is concerned, and the phase space is delineated for these sysyems. The triaxial distribution of velocities of objects of spherical systems and that of intermediate ones is investigated.\\After the Jeans theorem, the first integral of motion $\tilde{\Psi}$ ('phase density') is classified as a function of six independent first integrals of motion $\tilde{I}_i$ as a function of the phase-space variables as $\tilde{\Psi}=\tilde{\Psi}(\tilde{I}_1, \tilde{I}_2,..., \tilde{I_6})$. Each $\tilde{I}_i=const$ fixes $6$ $5$-dimensional hypersurfaces, which are moving within the phase space. Accordingly, the first integral $\tilde{\Psi}$ is constant in phase-space points moving with stars.\\ The density $\tilde{\rho}$ is determined after the Poisson equation, where the latter poses constraints not only on the phase density $\tilde{\Psi}$, which cannot therefore be assumed as arbitrary, but also on the density $\tilde{\rho}$, which cannot be assumed as as negative. The boundary conditions have to be satisfied at infinity, for which the standard cosmological principle has to be applied.\\ The Jeans theorem, furthermore, fixes the potential and the phase density as independent of time. When stationary potential are considered, orbits are classified according to the number of integrals $\tilde{J}$ needed for the description in the phase space, as $\tilde{\Psi}=\tilde{\Psi}(\tilde{J}_1, \tilde{J}_2,..., \tilde{J}_n)$, with $n\le5$.\\
A method similar to that of \cite{kuz53} is later found in \cite{mto07}; in particular, in the case of a plane galactic disk, after three isolating integrals of motion, the three invariants of motion are formulated as combinations of the 'elements of a Keplerian ellipse'.\\
In the case of Kuzmin-like potential, some integrals of motions are investigated in \cite{tvo01}.\\ 
In \cite{kuz87},  flattened stellar systems are outlined to be a complicated model in the case the phase density is provided with as a function of three integrals of motion, which
correspond to the triaxial local-velocity-distribution ellipsoid.\\
\section{Outlook\label{outlook}}
In \cite{Kormendy:2013nda}, the evolution of disk galaxies is investigated; the changes in the angular momentum are studied as due to the interactions between stars or of gas.\\
In \cite{Xu:2022ypi}, dark matter is presented to model the dumping.\\
In \cite{Vandervoort:2002vg}, the evolution of galaxies as far as the stationary oscillations are concerned is performed and analysed according to the analysis hinted in \cite{fpo84}.\\
In \cite{con1960}, a further integral of motion is found, of which the convergence is questioned about.\\
Isolating integrals of motion and non-isolating integrals of motion are discussed in \cite{lyn60} and \cite{lyn62}; in particular, in \cite{lyn62}, a distribution function is proposed to be built from isolating integrals only. For these purposes, the orbital phase angle is studied in \cite{fre75a} and \cite{fre75b}.\\
Non-analytical descriptions descending from the Kuzmin potential are presented in \cite{ken80}.\\
In \cite{kuz63} the very specific case of gravitational interaction within stars in clusters are studied. A host of detailed results are found. Among the findings, it is interesting to remark that the variation of the phase density was derived. The velocity dispersion and the mean dispersion are given.
\section{Perspectives\label{perspectives}}
In \cite{lyn62}, the Jeans theorem is applied to axially-symmetric stellar systems, and specified to self-gravitating systems. The space density is hypothesized to be a function of the gravitational potential and of the radial coordinate in cylindrical coordinates, form which an equation of the mass of the stellar objects per chosen density is obtained; from the solution, the velocity dispersion and the mean velocity are found, while the rotational velocity is shown to have to be determined after further assumptions. Indeed, the virial theorem is applied locally and the condition that the start should be not streaming but only rotating (where the latter is called a 'relaxation condition') is applied. A distribution function which allows one for the description of an observed
density variation is found.\\
In \cite{lyn60}, a spherical cluster of mass points is illustrated to be able to rotate without becoming oblate. The example of the Sun in the Galaxy is issued.\\
The numerical analysis of the central parts of galaxies has been shown to exhibit noisy oscillations \cite{nam00}.
\section{Remarks\label{remarks}}
The analyses of the observational evidences as far as the fitting algorithms are concerned is based on the comparison between the values of the virialised radii and on that of the virialised mass of stellar systems; more in detail, there exist severe discrepancies among the values needed for the fitting algorithms and those estimated from a mathematical point of view.\\
The definition of generalised Hamiltonian can be explored within the search of generalised potential, which can comprehend the (generalised) time-dependent oscillator. The resulting Hamiltonian systems can be shown to admit constants of motion; in particular, the constants of motions can be obtained after the conservation laws of the Hamiltonian system.\\
The main results of the present work are the new finding of new conservation laws of new constants of motion after the Lewis invariant; the framing of the construction within its proper cosmological implementation allows one to discover that the quantities entering the definitions of the constants of motion are the virialised radius and the virialiesd mass of the stellar system. More in particular, within the formalism here developped it is also possible to relate the quantities involved int he pulsed potentials to the mathematical ones.\\
The cosmological implementation here presented is one related to the age of non-Gaussianities; the formalism also straightforward applies to different ages of the Universe, at which the considered stellar system are estimated to be formed.\\
The modification of the behavior of the stellar system with respect to the new constants of motions is found to be ascribed to the phenomena modifying the CBE description.\\  
The paper is organised as follows.\\
In Section \ref{section1}, the main results relating constants of motion, the virial radius and the virial mass of stellar systems are scrutinised.\\
In Section \ref{section2}, the application of the virial theorem which defines the virialised radius (of a stellar system) is summarised.\\
In Section \ref{section3}, the physical implementation of the requests that lead to a  collisionless Boltzmann distribution function are recapitulated.\\
In Section \ref{section4}, oscillations of stellar systems are revisited.\\ 
In Section \ref{section5}, generalised Hamiltonians are introduced.\\
in Section \ref{section6}, in particular, the time-dependent oscillator potentials are enumerated.\\
In Section \ref{section7}, new estimations of the constants of motions after the Lewis-Ermakov-Leach invariant are found.\\
In Section \ref{cola}, the new conservation laws of the new constants of motions are stated. As results, the constants of motions are demonstrated to depend on the virialised mass and on the virialised radius of the stellar system. The new methods can be demonstrated to apply to the cases of the Ermakov-Lewis invariant and to that of the Ermakov-Lewis adiabatic invariant.\\
In Section \ref{section12}, the new constants of motion from the new conservation laws are implemented from a cosmological point of view.\\ 
In Section \ref{modcbe}, the modifications of the CBE hypotheses are debated. 
In Section \ref{remarks}, the new results are framed within the modern research guidelines; in particular, a comparison with the data-analysis techniques is implemented.

\end{document}